\documentclass[sigconf]{acmart}

\acmConference[Internetware 2026]{The 17th International Conference on Internetware}{18--20 July 2026}{Gold Coast, Australia}

\usepackage{listings}
\usepackage{xcolor}
\usepackage{graphicx}
\usepackage{amsmath}
\usepackage{multirow}
\definecolor{codegreen}{rgb}{0,0.6,0}
\definecolor{codegray}{rgb}{0.5,0.5,0.5}
\definecolor{codepurple}{rgb}{0.58,0,0.82}

\lstdefinelanguage{TypeScript}{
  keywords={public, private, protected, static, class, interface, extends, implements, constructor, function, return, void, number, string, boolean, if, else, true, false, null, undefined, throw, new},
  morecomment=[l]{//},
  morecomment=[s]{/*}{*/},
  morestring=[b]",
  morestring=[b]'
}

\newcommand{\tool}{HapCiD}

\newcommand{\find}[1]{%
  \par\smallskip\noindent\emph{#1}\par\smallskip
}

\title{HapCiD: Detecting API-related Compatibility Issues in OpenHarmony Apps}

\author{Daihang Chen}
\affiliation{%
  \institution{Beihang University}
  \city{Beijing}
  \country{China}}
\email{chendaihang@buaa.edu.cn}

\author{Yonghui Liu}
\authornote{Corresponding author.}
\affiliation{%
  \institution{Australian National University}
  \city{Canberra}
  \country{Australia}}
\email{Yonghui.Liu@anu.edu.au}

\author{Mingyi Zhou}
\affiliation{%
  \institution{Beihang University}
  \city{Beijing}
  \country{China}}
\email{zhoumingyi@buaa.edu.cn}

\author{Li Li}
\affiliation{%
  \institution{Beihang University}
  \city{Beijing}
  \country{China}}
\email{lilicoding@ieee.org}

\begin{document}

\begin{abstract}

OpenHarmony, an emerging open-source mobile platform, is rapidly gaining attention in the mobile development community. Its fast-evolving Software Development Kit (SDK) introduces numerous Application Programming Interfaces (APIs) to boost developer productivity but inevitably introduces compatibility challenges—an issue well known from platforms like Android. However, existing compatibility analysis tools are ineffective for OpenHarmony due to its newly introduced ArkTS programming language and the lack of a mature behavioral model to capture app execution semantics.
To bridge this gap, we present HapCiD, an open-source tool for automatically detecting API-related compatibility issues in OpenHarmony apps. Applied to 4,478 apps, HapCiD successfully detects 2,040 compatibility issues across 163 apps with 100\% accuracy. Based on these findings, we conduct a comprehensive study of compatibility issues, categorizing their types, examining existing mitigation strategies, and proposing future improvements. HapCiD and all experimental data are publicly available at: \url{https://github.com/ElouanDai/HapCiD}.

\end{abstract}



\maketitle

\renewcommand{\lstlistingname}{Code}

\section{Introduction}
\label{sec:introduction}


OpenHarmony is an open-source mobile operating system (OS) initially developed by Huawei and now maintained by the OpenAtom Foundation~\cite{openatomcon}.
As the operating system continually updates, OS providers periodically release new Software Development Kit (SDK) versions to enable developers to harness the full potential of a device ~\cite{li2016accessing}. 
These frequent updates, however, can lead to compatibility issues—particularly when applications depend on APIs that aren't supported across their specified system versions.

The consequences of API compatibility issues are well documented in Android, a mature mobile OS ecosystem. Incompatible API usage has caused severe functional and security issues, such as application crashes in widely used media streaming apps relying on deprecated audio APIs in Android API levels 21 and 23, or  potential security vulnerabilities in Notification API implementations allowing unauthorized notification access on Android 12 devices~\cite{exoplayercrash9704, android12notification}.

OpenHarmony is a newly designed mobile OS featuring multiple SDK versions, facing similar risks. 
Like Android, it lets developers pick which system versions the applications work on, eliminating compatibility issues. Developers could set both \textit{Compatible} and \textit{Compile} settings for each application, similar to Android's \textit{minSdkVersion} and \textit{targetSdkVersion}. These attributes inform users and app marketplaces about the SDK versions supported for the application.
Developers who set the \textit{Compile} SDK version higher than the actual version may inadvertently create \textit{backward compatibility issues}. Conversely, if the \textit{Compatible} version is lower than the minimum SDK version, it could potentially lead to \textit{forward compatibility issues}. 

Although various solutions have been explored to automate the identification of compatibility issues in Android apps~\cite{wei2016taming, huang2018understanding, yang2018android, li2018cid, kong2018automated, fazzini2019automated, haryono2020automatic, haryono2022androevolve}. For example, Wei et al.~\cite{wei2018understanding} developed a tool called FicFinder to automatically identify compatibility issues caused by Android fragmentation. However, the tool requires significant human effort to summarize rule-based API/context pairs from empirical studies. 
Huang et al.~\cite{huang2018understanding} proposed CIDER to primarily detect callback API compatibility issues. It requires heavy manual efforts to construct a graph-based model for each SDK version to capture control flow inconsistencies resulting from API evolution. 
However, existing tools for Android cannot be applied to OpenHarmony Apps, primarily due to the following reasons, which will be elaborated on in the background section: (1) \textbf{OpenHarmony utilizes a novel framework supported by a layered architecture}, and (2) \textbf{OpenHarmony apps are developed using a newly designed language called ArkTS.}
Therefore, there is a strong need to invent an automatic detection method for the compatibility issue of OpenHarmony Apps.

To fill this gap, in this work, we propose an approach called \tool{} for OpenHarmony Apps, which automates the detection of ArkTS API-related compatibility issues in OpenHarmony Apps. Specifically, we design an analysis framework to support the unique features of OpenHarmony Apps, the layered architecture, and the newly designed ArkTS programming language ArkTS. \tool{} automatically extracts OpenHarmony APIs from each SDK version and then models and characterizes the corresponding lifecycle for each API. This allows for the detection of potential compatibility issues arising from the APIs invoked in individual applications.

Based on our proposed tool, \textbf{1)} we investigate the compatibility issue of existing OpenHarmony apps, Similar to Android, from 4,478 OpenHarmony projects, we identify 2,040 compatibility issues, distributed across 163 different open source applications. We manually evaluate these detected issues and confirm that our tool achieves 100\% accuracy in detecting the API compatibility issue of OpenHarmony Apps. \textbf{2)} We then categorize the issues into two groups: backward and forward compatibility issues. Our results show that 1,144 of them are forward compatibility issues and 896 of them are backward compatibility issues, attributed to several major updates between API versions 10 and 20. Among them, we found that 48.5\% of the OpenHarmony applications use deprecated APIs. It shows that developers need to pay more attention to this, as OpenHarmony is a rapidly evolving operating system. \textbf{3)} Next, we collected and accessed the commit records of 44 open source repositories that have compatibility issues to investigate existing solutions. From the systematic analysis of these commits, we summarize the solutions to each category of API compatibility issues and provide case studies for them. \textbf{4)} From the commit record, we find that developers have not paid attention to compatibility issues in most cases. We notice that a large number of developers just try to modify the compatible SDK version of an application; thus, they do not accurately mitigate them, and the original problems persist even after the modification. Based on our findings, our method also offers the corresponding repair suggestions that can effectively address problems after they have been identified~\cite{liu2022autoupdate}.

Our contributions are shown below.
\begin{itemize}
    \item We introduce \tool, the tool designed specifically to detect compatibility issues in OpenHarmony applications. Through manual evaluation on an open source app dataset, \tool{} identified 1,144 backward compatibility issues and 896 forward compatibility issues, demonstrating 100\% detection accuracy.
    \item We provide a detailed analysis and summary of the compatibility issues detected across various OpenHarmony applications, offering targeted solutions based on case study evaluations.
    \item We present an overview of the evolution of OpenHarmony APIs to examine the prevalence of existing issues and solutions within current OpenHarmony applications. This investigation also explores the level of awareness and concern about API compatibility issues among developers, providing future improvements to solve them.
    \item We open source our proposed \tool{} and the collected dataset that has real-world OpenHarmony apps.

\end{itemize}

\begin{figure*}[t]
    \centering
    \includegraphics[width=\linewidth]{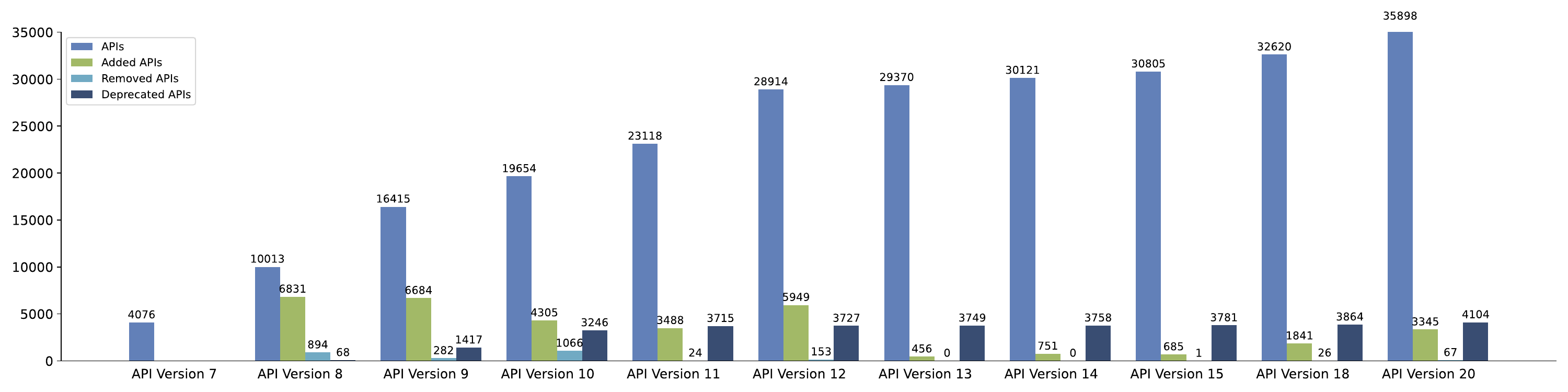}
    \caption{The statistics of API numbers of OpenHarmony versions}
    \Description{A line chart showing the growth in the number of OpenHarmony APIs across successive versions.}
    \label{fig:api_evolution}
\end{figure*}

\section{Background}

\subsection{OpenHarmony Versions and API Evolution}

Since its 2020 launch, OpenHarmony has undergone major evolution. Version 1.0 provided foundational API support for lightweight devices. Version 2.0 integrated Linux kernel for enhanced scalability and stability. Version 3.0 strengthened distributed architecture for cross-device collaboration. Version 4.0 added over 4,000 APIs, improved hardware adaptation, graphics, and distributed data management.
The 5.0 series marked significant progress: Version 5.0.0 (API 12) introduced enhanced application lifecycle management, ArkUI custom rendering, app cloning, and improved security and media support. OpenHarmony 6.0 (API 20) further improves ArkUI layouts, window management (including in-window text display), distributed data handling (via asset and asset group support), location services, and the input method framework. 

Prior to OpenHarmony v3.0 (API 7), application development relied primarily on Java. Since then, ArkTS has become the primary development language. Built on TypeScript, ArkTS integrates with the ArkUI framework to provide declarative UI and state management capabilities, enabling more concise and intuitive cross-platform development. \textbf{As ArkTS has superseded the Java version, this work focuses exclusively on ArkTS-based applications.}

\begin{table}[htbp]
\centering
\caption{The version information of major OpenHarmony versions.}
\resizebox{1\linewidth}{!}{
\begin{tabular}{|c|l|c|c|}
\hline
\textbf{NO.} & \textbf{OpenHarmony Version} & \textbf{API Version} & \textbf{Release Date} \\ \hline
1 & OpenHarmony-v3.0-LTS & 7 & 2021/09 \\ \hline
2 & OpenHarmony-v3.1-Release & 8 & 2022/03 \\ \hline
3 & OpenHarmony-v3.2-Release & 9 & 2023/04 \\ \hline
4 & OpenHarmony-v4.0-Release & 10 & 2023/10 \\ \hline
5 & OpenHarmony-v4.1-Release & 11 & 2024/03 \\ \hline
6 & OpenHarmony-v5.0.0-Release & 12 & 2024/09 \\ \hline
7 & OpenHarmony-v5.0.1-Release & 13 & 2024/11 \\ \hline
8 & OpenHarmony-v5.0.2-Release & 14 & 2025/01 \\ \hline
9 & OpenHarmony-v5.0.3-Release & 15 & 2025/03 \\ \hline
10 & OpenHarmony-v5.1.0-Release & 18 & 2025/05 \\ \hline
11 & OpenHarmony-v6.0-Release & 20 & 2025/09 \\ \hline
12 & OpenHarmony-v6.1-Release & 23 & 2026/03 \\ \hline
\end{tabular}}
\label{tab:openharmony_releases}
\end{table}

Table 1 shows the correspondence between each major OpenHarmony version, API level, and their respective release dates. We focus exclusively on the Release versions of these important major updates as representatives for API evolution analysis (therefore, intermediate or non-Release API levels such as 16, 17, and 19 are not included in the scope of this study). We choose to use the Release version of each major release as a representative sample and consider the ArkTS version of OpenHarmony 3.0 as the reference benchmark for studying the evolution of API versions. Although the OpenHarmony 6.1 Release version was released in March 2026, the official OpenHarmony community has stated that it plans to release the OpenHarmony 6.1 LTS version before June 30, 2026, as the new primary long-term maintenance and compatibility evaluation version~\cite{openharmony_v6_1_release}. Therefore, this paper only considers the versions up to OpenHarmony 6.0 Release (API 20) and earlier.


We focus specifically on the changes in the number of APIs from API version 8 to API version 18 to explore the potential API compatibility issues arising from updates in OpenHarmony versions. Using SDKs from long-term support and release versions as API version representatives, we parsed and statistically evaluated API changes.

Figure \ref{fig:api_evolution} displays the total number of APIs for each version from API 8 to API 20, including the number of API additions, removals, and deprecations. The total number of APIs has increased from 10,013 in API version 8 to 35,898 in API version 20, reflecting a total growth of 2.58 times across twelve versions. Our analysis includes the count of APIs marked with @deprecated in each OpenHarmony SDK version, which constitutes approximately 9.3\% of all API methods by API version 20. In addition, we have developed an API lifecycle model that facilitates querying the addition or removal of each API, allowing calculation of the net change in API numbers between consecutive versions. We observed that a substantial number of APIs were added during the transitions from API version 8 to 9 and from API version 11 to 12. Concurrently, as API versions evolved, a significant number of APIs were removed, especially in API version 10, where more than 1,000 APIs were removed compared to version 9. Without appropriate measures to protect against behavior that accesses unavailable APIs, applications could crash, negatively impacting user experience. Similarly, using deprecated APIs can also pose compatibility issues; these APIs might harbor security vulnerabilities, compromising application safety, and are therefore not recommended for use.


\begin{figure}[htbp]
    \centering
    \includegraphics[width=\linewidth]{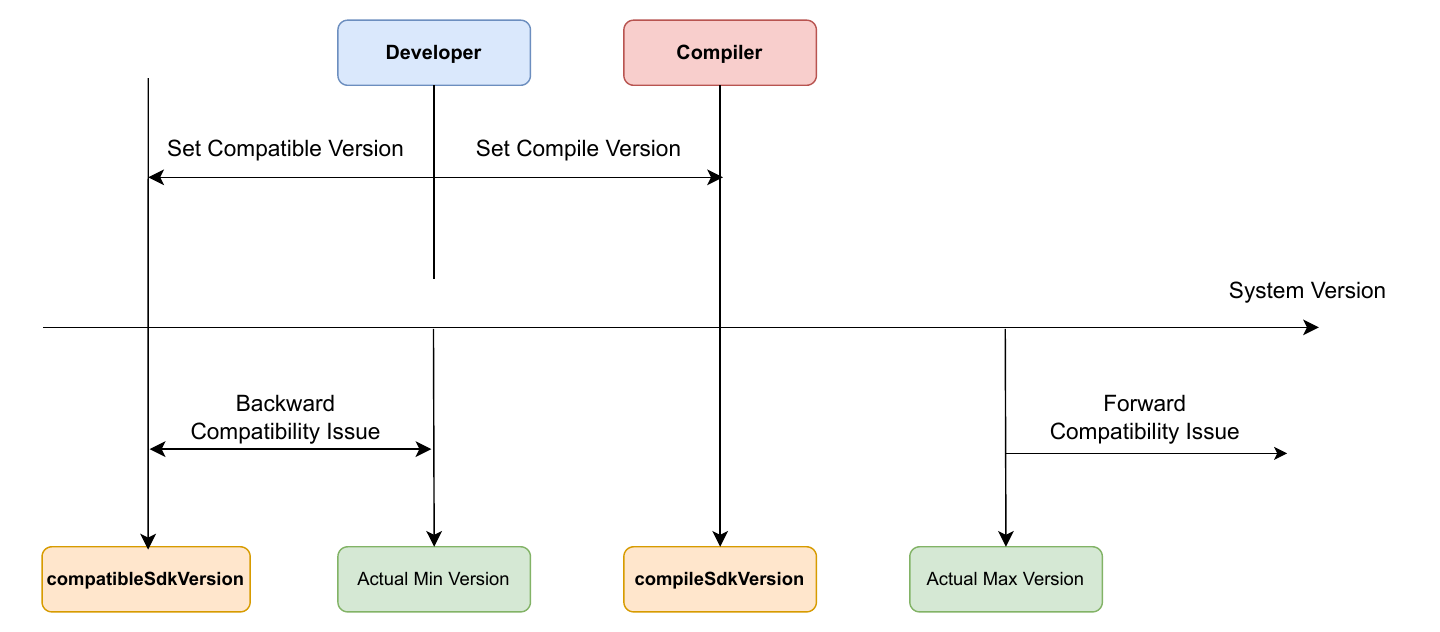}
    \caption{Examples of Forward and Backward Compatibility. 
    }
    \Description{Examples illustrating forward and backward API compatibility across OpenHarmony versions}
    \label{fig:compatibilityIssues}
\end{figure}

\subsection{API Compatibility Issues}

In the rapidly evolving landscape of mobile software development, API compatibility issues pose significant challenges for developers and users alike. As the OpenHarmony operating system continues to update its Software Development Kit (SDK), each new version introduces changes to the API set, including adjustments, removals, and additions of APIs to support emerging functionalities. Consequently, invoking APIs on devices incompatible with the SDK version can lead to compatibility issues, resulting in poor user experiences such as application installation failures or crashes when accessing certain features.

As shown in Figure~\ref{fig:compatibilityIssues}, the origins of API compatibility issues in OpenHarmony apps can be traced to the discrepancy between the device's SDK version and the version specified at compile time. Applications operate as expected when the device's OpenHarmony SDK version is equal to or higher than the compile-time version. 
\noindent\textbf{Backward Compatibility Issues} may arise when developers set the compatible version lower than the actual compatible SDK version, causing application crashes due to the invocation of inappropriate APIs on versions between the compatible and actual compatible versions. 
\noindent\textbf{Forward Compatibility Issues} occur when an application runs on versions higher than its maximum supported version, leading to failures when related APIs are called.

Existing research on Android API compatibility issues highlights that many mobile application developers struggle to accurately control the range of versions their applications use and often fail to respond to SDK version updates. This lack of proper version management significantly affects the user experience, as applications become unable to run on systems equipped with the appropriate SDK version. To address this challenge, a tool that can detect the correct SDK range for OpenHarmony applications is crucial for developers, enabling them to ensure smooth operation across various device configurations and SDK versions.

\subsection{Compatibility Issue Example in OpenHarmony Apps}

Listing~\ref{code:motivation} shows code snippet from an open-source app to demo the usage of feature, International~\cite{openharmony_international}. The Compatibility and Compilation settings for this app are set at 8 and 9, respectively. This suggests that the developer intends for the app to function optimally on an API Version 9 operating system while maintaining minimum compatibility with API Version 8. Unfortunately, within this app, the developer invokes API ``\textit{ohos.i18n.i18n.getSystemCountries(language: string)}'' and ``\textit{ohos.i18n.i18n.isSuggested(language: string, region?: string)}'', as illustrated in Line 3 and 8, which were introduced into package ``ohos.i18n.i18n'' in API Version 7 and were both moved into package ``ohos.i18n.system'' in API Version 9. Thus, if the app runs on a mobile device with an API Version 9 environment and invocations of these API are triggered, it will result in the app crashing. This is a typical case of a forward compatibility issue, where the developer inadvertently calls an API that is not compatible with the specified version and fails to implement versioning protection when the potentially problematic API is invoked. Such instances are not coincidental during our preliminary exploration. Therefore, it is crucial to employ a detection tool specifically designed to address API-induced compatibility issues arising from inappropriate OpenHarmony SDK version settings.

\begin{lstlisting}[label={code:motivation},language=TypeScript,caption={An example of Forward Compatibility Issue.}]
aboutToAppear() {
    Logger.info(TAG, `this.country = ${this.country}`)
    this.countryIds = i18n.getSystemCountries(this.localLanguage)
    Logger.info(TAG, `systemCountryIds = ${JSON.stringify(this.countryIds)}`)
    this.countryIds.forEach(id => {
        let country = i18n.getDisplayCountry(id, this.localLanguage)
        this.countries.push({ key: country, value: '' })
        if (i18n.isSuggested(this.localLanguage, id)) {
            this.suggestIds.push(id)
            this.suggestCounties.push({ key: country, value: '' })
        }
    })
}
\end{lstlisting}
\section{Approach}
\label{sec:approach}

\begin{figure*}[t]
    \centering
    \includegraphics[width=0.8\linewidth]{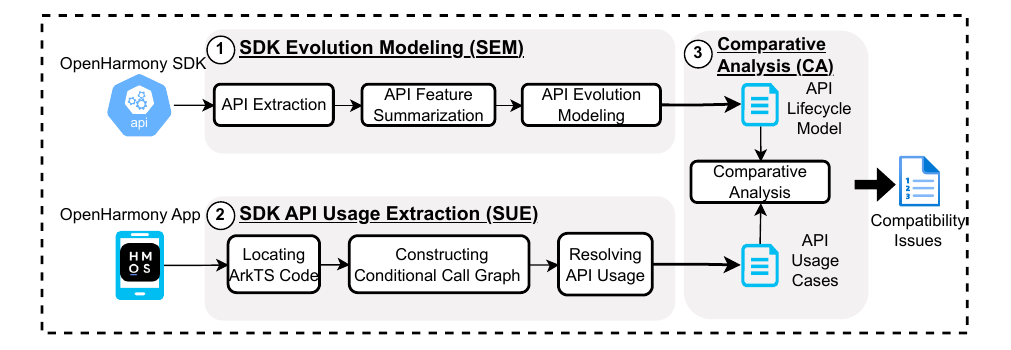}
    \caption{Overview of \tool.}
    \Description{The HapCiD workflow, from OpenHarmony SDK API extraction and lifecycle modeling to compatibility issue detection in applications.}
    \label{fig:overview}
\end{figure*}


\tool{} aims to detect API compatibility issues in OpenHarmony applications by comparing the SDK API model with its actual usage. To achieve this, \tool{} first builds a comprehensive model for the previously unexplored SDK usage. It then identifies usage instances within the targeted apps, which is challenging due to OpenHarmony's novel development language. Finally, \tool{} performs a comparative analysis to detect abnormal utilization of the corresponding SDK APIs.

\subsection{SDK Evolution Modeling (SEM)}
The aim of the SEM module is to create a dependable and reusable framework that allows for the integration and querying of method and field API lifecycle information by any tool or method that requires it.

To accomplish the SDK Evolution Modeling, SEM is divided into the following two steps:

\begin{lstlisting}[label={code:animator},caption={Example API Declaration}]
export default class Animator {
  /**
   * Create an animator object for custom animation.
   * @param { AnimatorOptions } options - Options.
   * @returns { AnimatorResult } animator result
   * @syscap SystemCapability.ArkUI.ArkUI.Full
   * @since 6
   * @deprecated since 9
   * @useinstead ohos.animator.create
   */
  static createAnimator(options: AnimatorOptions): AnimatorResult;
}
\end{lstlisting}

\textbf{API Extraction}. From the interface\_sdk-js\cite{api_parser_tool} repository on the official OpenHarmony organization on the Gitee platform, the source code and API declaration files for historical versions of OpenHarmony are obtained. The APIs declared in the Long-Term Support or release versions of the historical OpenHarmony SDKs represent all the APIs included in a particular version of an API. Building on the basic API parsing tool, collect\_API, functionality is extended and enhanced to extract API information. The specific implementation method involves using Node.js's fs module to read .d.ts files in the project, which contain OpenHarmony's API definitions; parsing these definition files to convert TypeScript code into an abstract syntax tree; traversing the AST to precisely identify which APIs are declared, including method names, parameter lists, parameter types, and return types, with special attention given to annotations indicating APIs that have been deprecated or modified in specific versions; and collecting and summarizing all API information to generate corresponding API reports for each OpenHarmony SDK version.
Below, using an API declared in the @ohos.animator.d.ts file from the OpenHarmony-v5.0-Release version SDK that is used to create animations in OpenHarmony applications as an example, the key API information and the extraction process focused on in the API extraction phase are detailed.

\begin{lstlisting}[label={code:generic_varargs},caption={Example API Syntax Features}]
//@ohos.file.photoAccessHelper.d.ts
declare namespace photoAccessHelper {
  interface MediaAssetDataHandler<T> {
    onDataPrepared(data: T, map?: Map<string, string>): void;
  }
}
//@ohos.account.appAccount.d.ts
declare namespace appAccount {
  interface AppAccountManager {
    addAccount(name: string, extraInfo?: string): Promise<void>;
  }
}
// Example usage of MediaAssetDataHandler
class PhotoHandler implements photoAccessHelper.MediaAssetDataHandler<string> {
  onDataPrepared(data: string): void {
    console.log(`Received data: ${data}`);
  }
}
// Example usage of AppAccountManager
const accountManager = new appAccount.AppAccountManager();
accountManager.addAccount("user1");
accountManager.addAccount("user2", "extraInfo2");
\end{lstlisting}

\textbf{API Feature Summarization}.
Accurately capturing API lifecycles in ArkTS demands careful handling of advanced features like generics and varargs, which complicate signature matching in application code. These syntactic complexities make building a precise lifecycle model challenging, as detailed below.

\begin{itemize}
\item \textit{Generic Type.} ArkTS inherits TypeScript's basic syntax. Generic functions offer flexibility by adapting to various data types while maintaining type safety. Generics empower SDK APIs to create functions that work with various data types. While Generic Types simplify maintenance and reduce code duplication, they can complicate the modeling of an API's lifecycle, making it more challenging to determine the API's lifespan accurately.
For instance, consider the method "\textit{onDataPrepared(data: T, map?: Map<string, string>): void}" in the "\textit{MediaAssetDataHandler}" interface (line 4 in Listing~\ref{code:generic_varargs}), which incorporates a generic type (T). When matching this signature with the application code, it may not always be straightforward to identify usage instances, as demonstrated by line 15 in Listing~\ref{code:generic_varargs}.

\item \textit{Varargs.} Like API methods with generic types, certain API methods employ variable arguments (varargs), which can pose challenges when trying to detect their usage in OpenHarmony apps. For example, consider the API method "\textit{addAccount(name: string, extraInfo?: string): Promise<void>}" within the "\textit{AppAccountManager}" interface (line 10 in Listing~\ref{code:generic_varargs}), which includes a varargs parameter. The actual usage of this API might involve "\textit{addAccount(string, string)}" or "\textit{addAccount(string, string, string)}" (lines 21 and 22 in Listing~\ref{code:generic_varargs}), and in any scenario, it will exhibit distinct syntactical characteristics compared to the method signature.
\end{itemize}

To address the potential inaccuracies and omissions resulting from the above two challenges, SEM incorporates an analysis phase for those extracted API Information. During this phase, SEM enriches the gathered API signatures with additional data concerning the possible concrete parameters and their corresponding types. This augmentation enhances the accuracy and completeness of the information collected.

\textbf{API Evolution Modeling}.  Following the augmentation of APIs, SEM processes consecutive versions of the APIs to determine if the current API has been encountered previously. Through this method, we pinpoint the versions in which each API was introduced and subsequently removed. After analyzing all versions, we successfully construct a model representing the lifecycle of APIs. Engaging with this model enables the identification of the lifecycle for any particular API. 
For example, the lifecycle for the \textit{getSystemCountries(language: string)} API within the \textit{ohos.i18n} namespace mentioned in Listing \ref{code:motivation} starts from API Version 7 and continues through to API Version 8.

As SEM's extraction process is entirely automated, any updates to the API's lifecycle extracted by SEM can be effortlessly accommodated, even in the event of updates to the OpenHarmony SDK. It is important to emphasize that the ALM module, along with its associated query interface, is entirely decoupled from the implementation of \tool. This independence makes it easily adaptable to other methods and tools. The present \tool{} implementation has already incorporated a query interface to facilitate such adaptability.

\subsection{SDK API Usage Extraction (SUE)}
In this module, our objective is to identify all APIs invoked by the scrutinized application and extract those APIs that meet the detection criteria as the final output. It is essential to emphasize that SUE requires access to the source code of the application. To perform API extraction, SUE is divided into the following three sequential steps:

\textbf{Locating ArkTS Code and Converting to TS Code.} When building OpenHarmony applications, developers distribute module functionalities across multiple files using the ArkTS programming language. We use the ArkTS2TS tool to identify all files with the extension '.ets' and convert them into semantically equivalent ts files. The converted results, along with other ts code from the application, are then loaded into a static analysis tool called ArkAnalyzer\cite{chen2025arkanalyzer}(a tool specifically introduced by the OpenHarmony team for performing static analysis on OpenHarmony applications) and linked with the corresponding version of the OpenHarmony SDK.

\textbf{Constructing Condition Call Graph.}
We can identify APIs that may cause compatibility issues by checking whether the detected APIs fall within the version range specified by the application developer. However, this approach may lead to false positives. This is because some developers adopt the API protection strategy described below to ensure that API calls are made only when the device version is within the required range.

\textit{API Protection strategy} This strategy aims to ensure consistent functionality across various device versions and prevent errors by selectively invoking different APIs based on the current SDK version. OpenHarmony's SDK provides the ``deviceInfo.sdkAPIVersion'' API, which allows developers to obtain the API version of the device running the application. Based on this, developers can decide whether to call a specific API and whether the system needs to throw an exception to alert the user, thereby reducing potential risks caused by API compatibility issues.

Listing \ref{code:api_protection} demonstrates a real-world implementation of API protection in the ohos-MPChart application from the OpenHarmony-SIG organization. The application utilizes potentially version-sensitive APIs like clip() and fill() from the ArkUI.CanvasRenderer module which is only suitable for API version 8 or higher. Thus, the developer employs the getSDKInt() method to determine the API version of the device on which it is running. This method provides a basis for determining the support for the clip() method, protecting against compatibility issues with APIs during application execution. For instance, within the ensureClipPathSupported() function, if the API version is found to be too low, the system will throw an exception.

\begin{lstlisting}[language=TypeScript, label={code:api_protection}, caption={An Example of API Protection Mechanisms}]
// (a) Example 1 of API Protection
public static getSDKInt(): number {
  return deviceInfo.sdkAPIVersion;
}

private isClipPathSupported(): boolean {
  return Utils.getSDKInt() >= 8;
}

private ensureClipPathSupported(): void {
  if (Utils.getSDKInt() < 8) {
    throw new Error("Fill-drawables not (yet) supported below API level 8, " +
    "this code was run on API level " + Utils.getSDKInt() + ".");
  }
}

// (b) Example 2 of API Protection
public fillRect(c: canvasRenderingContext2D, ...): void {
  switch (this.mType) {
    ...
    case FillType.COLOR:
      if (this.isClipPathSupported()) {
        ...
        c.clip();
        ...
      }
    ...
    case FillType.DRAWABLE:
      this.ensureClipPathSupported();
      ...
      c.clip();
      ...
  }
  ...
}
\end{lstlisting}

The fillRect() function is an integral part of the main source code logic of the application, which effectively utilizes the two previously mentioned functions for API protection. This strategy determines whether to invoke a specific API or whether to throw an exception. In response to this issue, \tool{} also employs call graph methods to identify if there are any API protection issues within the application's source code. It has successfully detected the API protection behavior in this case, thereby preventing the generation of API compatibility issue alerts.

Existing research indicates that there are no cases in the Android domain where potential compatibility issues with APIs are preemptively safeguarded by throwing exceptions via functions such as ensureClipPathSupported(). Therefore, this finding within OpenHarmony applications can be considered a unique feature associated with applications developed using the ArkTS language, demonstrating the importance of the analysis as a typical case. 

To accurately identify protected APIs, we construct conditional call graphs to model API invocations under version-specific conditions. This involves analyzing the application’s control flow and data dependencies to determine whether API calls are safeguarded by version checks. Through a specialized analysis in the SUE module, we ensure APIs are invoked in appropriate system environments, enabling precise detection of protected APIs and reducing false positives.

\textbf{Resolve API Usage.}
SUE determines API usage by assessing whether all methods called and properties accessed by the application are OpenHarmony APIs. We focus on all APIs that have appeared across various versions of OpenHarmony. Upon detecting an OpenHarmony API, we generate an output that provides detailed information about the API, including its call pathway and any other relevant information.

\subsection{Comparative Analysis (CA)}

With the OpenHarmony API lifecycle model obtained from SEM and all API calls in the app under detection derived from SUE, CA's role is to correlate these two sets of information. Each API discerned by SUE can be referred to in the lifecycle model. Upon retrieving the respective lifecycle, CA compares it with the Compile and Compatible set by the developers. APIs that do not comply with the lifecycle are reported as compatibility issues. This approach allows us to generate dependable compatibility issue reports and effectively pinpoint potential compatibility issue APIs based on their paths. More specifically, the produced reports enumerate all the APIs invoked by the app with compatibility issues, detailing the lifecycle of these APIs as well as every method that has called upon these problematic APIs.

\section{Evaluation}

 To evaluate whether \tool{} can effectively detect compatibility issues in OpenHarmony applications, we aim to address the following four key research questions.

 \begin{table*}[tbp]
\centering
\caption{Distribution of Applications by Source and Compile API Version.}
\label{tab:applications}
\footnotesize 
\resizebox{\textwidth}{!}{ 
\begin{tabular}{|l|c|c|c|c|c|c|c|c|c|c|c|c|c|c|c|c|c|}
\hline
\textbf{Source} & \textbf{App Repos} & \textbf{Apps} & \textbf{8} & \textbf{9} & \textbf{10} & \textbf{11} & \textbf{12} & \textbf{13} & \textbf{14} & \textbf{15} & \textbf{16} & \textbf{17} & \textbf{18} & \textbf{19} & \textbf{20} & \textbf{Error} \\ \hline
OpenHarmony & 48 & 2572 & 135 & 170 & 94 & 59 & 124 & 4 & 186 & 27 & 9 & 0 & 36 & 2 & 1724 & 2 \\ \hline
OpenHarmony-SIG & 255 & 581 & 89 & 32 & 92 & 15 & 339 & 13 & 0 & 0 & 0 & 0 & 0 & 0 & 0 & 1 \\ \hline
OpenHarmony-TPC & 44 & 265 & 10 & 42 & 136 & 1 & 68 & 8 & 0 & 0 & 0 & 0 & 0 & 0 & 0 & 0 \\ \hline
Other Sources on Gitee & 52 & 336 & 3 & 44 & 35 & 8 & 166 & 7 & 3 & 5 & 13 & 28 & 5 & 18 & 0 & 1 \\ \hline
Github  & 89 & 724 & 226 & 258 & 123 & 18 & 64 & 9 & 3 & 15 & 1 & 2 & 2 & 0 & 3 & 0 \\ \hline
\textbf{Total} & 488 & 4478 & 463 & 546 & 480 & 101 & 761 & 41 & 192 & 47 & 23 & 30 & 43 & 20 & 1727 & 4 \\ \hline
\end{tabular}}
\end{table*}

\begin{itemize}

\item \textbf{RQ1:} How effective is \tool{} in detecting compatibility issues in OpenHarmony applications?

\item \textbf{RQ2:} What compatibility issues do OpenHarmony applications have?


\item \textbf{RQ3:} What practice do developers adopt on those compatibility issues?

\end{itemize}

\noindent\textbf{Data Collection:}
To our knowledge, no publicly available dataset exists for OpenHarmony applications. We curated a comprehensive dataset from Gitee and GitHub to (1) evaluate HapCiD’s detection capability and (2) analyze developer practices for addressing API compatibility issues.

As of March 2026, we collected 4,478 successfully compiled OpenHarmony applications written in ArkTS from 488 repositories. These applications use APIs ranging from versions 8 to 20, covering the full spectrum from the earliest to the latest API versions available for ArkTS-based development. Table~\ref{tab:applications} presents the distribution of applications by source and compile API version.

The applications are sourced primarily from three key organizations (OpenHarmony~\cite{openharmony_project}, OpenHarmony-SIG~\cite{openharmony_sig}, and openharmony-TPC~\cite{openharmony_tpc} ) in the OpenHarmony community, with additional repositories from GitHub and other repositories on Gitee to enhance coverage.

\subsection{RQ1 - Assessing \tool's Effectiveness}

To assess the effectiveness of \tool{} in detecting API compatibility issues, this study conducted an analysis on a dataset of 4,478 open source applications collected from the Gitee and GitHub platforms. The analysis detected 1,144 backward compatibility issues in 89 different applications and 896 forward compatibility issues in 87 applications.

To ensure the reliability and reproducibility of our findings, we performed a rigorous manual verification process with a clear, auditable protocol:

Ten researchers participated in the verification, divided into two independent groups of five people each. One group was responsible for verifying all forward compatibility issues, while the other group verified all backward compatibility issues. 

The verification criteria for each reported issue were as follows:
\begin{itemize}
    \item The application must be successfully installed and launched on the target OpenHarmony device.
    \item The specific API flagged by \tool{} as causing a compatibility issue must be triggered during normal usage of the corresponding functionality.
    \item We checked whether the functionality worked correctly or resulted in a crash / abnormal behavior due to the incompatible API.
\end{itemize}

All verifications were conducted on official OpenHarmony emulators and physical devices running the exact SDK versions reported in each application's \texttt{build-profile.json5} (compileSdkVersion and compatibleSdkVersion). Each issue was independently reviewed by at least two researchers within its group. 

After thorough verification, all 2,040 compatibility issues reported by \tool{} were confirmed to be true positives: the flagged APIs indeed caused either application crashes or malfunction of the corresponding features when the incompatible SDK version was used.



\begin{table}[htbp]
\centering
\caption{Results of open source applications detected by \tool.}
\label{tab:compatibility_issues}
\resizebox{\columnwidth}{!}{ 
\begin{tabular}{|c|l|c|c|c|c|}
\hline
\multirow{2}{*}{\textbf{No.}} & \multirow{2}{*}{\textbf{App Name}} & \multicolumn{2}{c|}{\textbf{Developer Settings}} & \multicolumn{2}{c|}{\textbf{Compatibility Issues}} \\ \cline{3-6} 
 &  & \textbf{Compile} & \textbf{Compatible} & \textbf{Forward} & \textbf{Backward} \\ \hline
1 & relationalStore & 20 & 9 & 0 & 166 \\ \hline
2 & Pasteboard & 20 & 9 & 0 & 122 \\ \hline
4 & photos & 9 & 9 & 84 & 0 \\ \hline
5 & preferencesEtstest & 20 & 9 & 0 & 64 \\ \hline
6 & InputMethodWind & 20 & 12 & 0 & 61 \\ \hline
7 & taskpool\_lib\_standard2 & 20 & 10 & 0 & 58 \\ \hline
8 & filemanager & 9 & 9 & 50 & 0 \\ \hline
9 & openharmony-learning & 10 & 10 & 44 & 3 \\ \hline
10 & dataObjectStagetest & 20 & 9 & 0 & 43 \\ \hline
11 & analogclock & 11 & 9 & 2 & 32 \\ \hline
12 & ipc\_stagetest & 20 & 11 & 0 & 32 \\ \hline
13 & FaceRecognition & 8 & 8 & 9 & 23 \\ \hline
14 & InputMethodEngine & 20 & 14 & 0 & 30 \\ \hline
15 & MediaCollections & 9 & 9 & 26 & 0 \\ \hline
16 & Recorder & 9 & 9 & 24 & 0 \\ \hline
17 & CameraView & 9 & 9 & 24 & 0 \\ \hline
18 & MultiMedia & 9 & 9 & 23 & 0 \\ \hline
\end{tabular}}
\end{table}

Table ~\ref{tab:compatibility_issues} lists the applications with more than 20 detected compatibility issues. The third and fourth columns display the range of developer version settings for the applications. In the OpenHarmony application configuration files, developers need to configure two properties: the compile SDK version (compileSdkVersion) and the compatible SDK version (compatibleSdkVersion). The compile SDK version is the SDK version that the code depends on during compilation, specifying the API library version used when building the application. Applications compiled with this version can access all the new APIs introduced in that SDK version. The compatible SDK version is used to declare the minimum SDK version with which the application is compatible, typically set to ensure the application can run on older SDK versions without issues. The fifth and sixth columns show the compatibility issues identified by \tool. Most backward compatibility issues were found in applications with a compatible version of 9. Out of the 4,478 applications, 163 had compatibility issues, accounting for 3.6\% of the total number of applications.

\begin{table}[htbp]
\centering
\caption{Representative APIs with Compatibility Issues.}
\label{tab:api_compatibility}
\small 
\resizebox{\columnwidth}{!}{ 
\begin{tabular}{|c|l|c|c|}
\hline
\textbf{No.} & \textbf{API} & \textbf{Lifecycle} & \textbf{Count} \\ \hline
\multicolumn{4}{|c|}{\textbf{Backward Compatibility}} \\ \hline
1 & ohos.data.relationalStore.RdbStore.createTransaction & 14+ & 53 \\ \hline
2 & ohos.pasteboard.SystemPasteboard.hasDataSync & 11+ & 28 \\ \hline
3 & ohos.data.relationalStore.RdbStore.batchInsertWithConflictResolution & 18+ & 23 \\ \hline
4 & ohos.pasteboard.SystemPasteboard.getMimeTypes & 14+ & 22 \\ \hline
5 & ohos.pasteboard.SystemPasteboard.getDataWithProgress & 15+ & 20 \\ \hline
6 & ohos.taskpool.SequenceRunner.execute & 11+ & 20 \\ \hline
7 & ohos.data.unifiedDataChannel.UnifiedData.constructor & 10+ & 18 \\ \hline
8 & ohos.rpc.MessageSequence.writeArrayBuffer & 12+ & 18 \\ \hline
9 & ohos.data.preferences.Preferences.getSync & 10+ & 16 \\ \hline
10 & ohos.data.unifiedDataChannel.UnifiedData.addRecord & 10+ & 16 \\ \hline
\multicolumn{4}{|c|}{\textbf{Forward Compatibility}} \\ \hline
1 & ohos.multimedia.mediaLibrary.mediaLibrary.getMediaLibrary & [6,11] & 86 \\ \hline
2 & ohos.multimedia.mediaLibrary.FileAsset.title & [7,11] & 71 \\ \hline
3 & ohos.multimedia.mediaLibrary.FileAsset.mediaType & [8,11] & 62 \\ \hline
4 & ohos.i18n.i18n.getSystemLanguages & [7,8] & 8 \\ \hline
\end{tabular}}
\end{table}

Table~\ref{tab:api_compatibility} presents the ten APIs most frequently associated with backward compatibility issues and the four most representative APIs causing forward compatibility issues, as forward compatibility problems typically arise from the removal of entire API packages. Most backward compatibility issues stem from new APIs introduced in API version 10 and later, which developers often fail to recognize as unavailable in earlier versions. Additionally, forward compatibility issues are predominantly caused by APIs in the ohos.multimedia.mediaLibrary module. This module was introduced in API version 6, deprecated starting from API version 9, and removed in API version 12, leading to compatibility issues when applications run on devices with higher API versions.


In summary, in the open source application dataset mentioned above, \tool{} identified 2,040 compatibility issues, including 1,144 forward compatibility issues and 896 backward compatibility issues, distributed in 163 different open source applications. We reviewed all compatibility issues caused by different APIs, manually located the problem modules in each application, reviewed the relevant code snippets, and ran the applications on OpenHarmony devices to trigger these suspect APIs to confirm whether they cause compatibility issues or lead to application crashes. All compatibility issues have been manually verified and confirmed as accurate, with each issue being reproducible on OpenHarmony devices.

\find{\textbf{Answers to RQ1:}
\tool{} demonstrates high effectiveness in detecting API compatibility issues within OpenHarmony open-source applications. By leveraging \tool{}, we can precisely identify APIs causing compatibility problems, ensuring accurate detection across the analyzed projects.
}

\subsection{RQ2 - Examining compatibility issue variations}
With proposed SDK API Evolution Modeling, compatibility issues fall into three categories:

\subsubsection{API Introduction}
This is the most prevalent type detected. It occurs when developers invoke APIs without verifying their existence in the compile SDK version or compatible SDK version, often leading to runtime errors, performance issues, or functional/security problems.

\textbf{Case Study:} ohos.pasteboard.SystemPasteboard.hasDataSync.

\tool{} identified a backward compatibility issue related to this API. The API was first introduced in OpenHarmony SDK API version 11. Applications using this API set the compatible SDK version to 9, causing compatibility issues when users run the application on OpenHarmony mobile devices with API version 9 or 10. For example, in the Pasteboard application, the results \tool{} show that this API is called within the PasteboardGetMimeTypesTest() function. Following this discovery, we ran the application using this API on an OpenHarmony mobile device with API version 10, resulting in an application crash. Subsequently, we extracted error information to confirm that the crash was due to an API compatibility issue.

\textbf{Repair Suggestions:} Considering the frequent addition of APIs in existing applications, we advocate two remedial methods. The first is to adjust the application's compatible SDK version from a lower version to one that matches the considered API. The second method involves selecting an alternative API based on the SDK version of the user's device. After applying the suggested fixes, we re-evaluated the applications with \tool. Modifying the compatible SDK version eliminated the alerts for this API compatibility issue, and applications using API protection strategy to safeguard backward compatibility also showed no detectable issues.

\subsubsection{API Deprecation}
This issue occurs when older versions of APIs are marked as deprecated in newer versions but are not immediately removed. Although these APIs remain usable, they are not recommended for use in the official documentation, and developers are encouraged to use newly introduced APIs instead. In fact, API deprecation is very common between OpenHarmony versions, leading to a significant number of applications in our dataset using deprecated APIs. Furthermore, there are numerous instances of entire modules being moved or deprecated in the history of OpenHarmony's API updates. Despite these APIs still being usable, we hope that developers will pay attention to the potential risks of compatibility issues they pose.

\begin{table}[htbp]
\centering
\caption{Usage of Deprecated APIs in Applications}
\label{tab:deprecated_apis}
\small 
\resizebox{\columnwidth}{!}{ 
\begin{tabular}{|l|c|c|c|}
\hline
\textbf{Source} & \textbf{Apps} & \textbf{Deprecated API Apps} & \textbf{Deprecated APIs} \\ \hline
OpenHarmony & 2572 & 1237 & 27247 \\ \hline
OpenHarmony-SIG & 581 & 156 & 1694 \\ \hline
OpenHarmony-TPC & 265 & 157 & 1152 \\ \hline
Other sources on Gitee & 336 & 109 & 990 \\ \hline
Github & 724 & 514 & 5904 \\ \hline
\textbf{Total} & 4478 & 2173 & 36987 \\ \hline
\end{tabular}}
\end{table}

Table ~\ref{tab:deprecated_apis} shows the statistics on the usage of deprecated APIs within the open source OpenHarmony application dataset. We found that 48.5\% of the applications use deprecated APIs. Within the main OpenHarmony organization of 2,572 applications, each application uses an average of 10.59 deprecated API instances, highlighting the need for developers to pay more attention.

\textbf{Case Study:} ohos.multimedia.mediaLibrary.

The media library management module was introduced in API version 6 and includes eight APIs to manage media assets. This module was deprecated starting from API version 9 and completely removed in API version 12. In API version 9, two of its APIs were replaced by updated ones in the ohos.file.photoAccessHelper module, while the remaining six APIs were left without direct alternatives. Additionally, one of the replacement APIs was also deprecated in API version 11 and was subsequently replaced by a new API within the same module.

Among the 896 forward compatibility issues detected in RQ1, 832 originate from APIs in the media library management module which has been removed, accounting for 92.9\% of the total forward compatibility issues. This significant proportion underscores the substantial compatibility risks associated with using deprecated APIs, highlighting the need for timely detection and resolution to prevent issues as API versions evolve.

\textbf{Repair Suggestions:} While these deprecated APIs may not cause severe compatibility issues or directly lead to application crashes, resolving these issues is crucial for developers since deprecated APIs might be removed in future SDK updates. In such cases, using the previously mentioned technique of obtaining API versions to safeguard APIs is often more effective than directly modifying the compatible SDK version, as the code can still run correctly under previous SDK versions. After implementing protection using the API safeguard mechanism, we are no longer able to identify similar compatibility issues.

\subsubsection{API Removal}
This type of forward compatibility issue occurs when older version APIs are removed from newly released SDKs, resulting in incompatibility between applications that use old versions and systems that run new SDKs.

\textbf{Case Study:} ohos.i18n.

The application "International" experienced six forward compatibility issues due to the deletion of APIs of the old version in a new version. Since API 9, some APIs from the ohos.i18n.i18n class within the ohos.i18n package were migrated to the ohos.i18n.System in the ohos.i18n package. This migration caused the application to malfunction on devices installed with API 9 or higher version SDKs, leading to severe compatibility issues. API calls for the compatibility problem were located in the IDE. When this application ran on a device emulator with API 9 installed, opening the page containing the problematic API calls caused the application to crash. This type of compatibility issue is difficult to detect and resolve during the development process because no corresponding alerts appear in the IDE linked to the old version SDK. This illustrates how potential compatibility problems can significantly affect the user experience when using the application and can also pose certain security risks.

\textbf{Repair Suggestions:} In this case, compatibility issues might cause a "method not found" error, leading to application crashes. Adjusting the application's compatible SDK version settings would limit the applicable SDK versions for the application, which is not the best solution. Instead, we recommend using an API protection mechanism to safeguard old-version APIs and employ custom functions or updated APIs for newer versions. After applying the suggested fixes to all affected open-source applications, we re-tested them with \tool. The results showed that modifiable compatibility issues were no longer detected, and all getAPIVersion-protected APIs were correctly identified as protected.

\find{\textbf{Answers to RQ2:}  
We provide a taxonomy of compatibility issues in OpenHarmony applications, categorizing them into API introduction, deprecation, and removal, and propose targeted fixes based on our findings and illustrative cases.
}

\subsection{RQ3 - Compatibility Issues While Developing}

To study how developers perceive and handle API compatibility issues in OpenHarmony applications, we analyzed commit histories of 44 repositories containing the 163 applications with issues found in RQ1. This allowed us to observe how these issues evolved over time and whether developers attempted to fix them.

\textbf{Setup:} For repositories officially managed within the OpenHarmony community, we analyzed all tagged versions with \tool{} and compared adjacent versions to track issue resolution. For repositories maintained by individual developers or third parties, which typically lack standardized tags, we selected 20 historical commits using equidistant sampling (for example, from the 413 commits in "harmony-utils" we sampled one every 20 commits).


\textbf{Results:} We tracked each application's evolution by comparing adjacent versions. Whenever compatibility patterns changed—new issues appeared or existing ones disappeared—we manually inspected the corresponding commits to understand the developer actions, and summarized the representative cases in Table~\ref{tab:commits}.

\begin{table}[htbp]
\centering
\caption{Compatibility Issues Fixed While Developing.}
\label{tab:commits}
\resizebox{1\linewidth}{!}{
\begin{tabular}{|c|c|c|c|c|}
\hline
\textbf{Repository} & \textbf{Commit} & \textbf{$\Delta$ Issues} & \textbf{Changes} \\ \hline
app\_samples & 97db735 & -20 & compatibleSdkVersion 6 to 9 \\ \hline
applications\_contacts & 818f173 & -16 & compatibleSdkVersion 9 to 10 \\ \hline
app\_samples & b6e58b5 & -13 & compatibleSdkVersion 9 to 10 \\ \hline
security\_privacy\_center & 9d87890 & -5 & compatibleSdkVersion 10 to 11 \\ \hline
applications\_print\_spooler & 5e330ef & -4 & compatibleSdkVersion 9 to 10 \\ \hline
applications\_contacts & f0065631 & +1 & compatibleSdkVersion 10 to 9 \\ \hline
\end{tabular}}
\end{table}


\subsubsection{Developer Awareness and Practices} 

\noindent\textbf{Troubling Awareness on Compatibility Issues.}
Our results show that most repositories ignore compatibility issues: once present, they persist unchanged across versions. For example, applications\_filepicker repeatedly uses the API 10 interface ``ohos.file.fs.Filter'' with compatibleSdkVersion 9, and the issue remains unresolved in 14 versions, revealing a fundamental lack of awareness.

\textbf{Successful Actions on Compatibility Issues.} Some developers promptly identify and fix compatibility problems by adjusting compileSdkVersion and compatibleSdkVersion. For instance, in one commit, the applications\_print\_spooler application in the main OpenHarmony organization raised compatibleSdkVersion from 9 to 10, which resolved issues introduced by APIs in the ohos.net.mdns package at API 10. These cases show that at least a subset of developers recognize API compatibility issues and can address them effectively.

\textbf{Failed action on Compatibility Issues.} However, we also found that a small number of developers do not fully understand API compatibility issues. They attempt to modify the compatibleSdkVersion, but the original compatibility issues still persist. Additionally, we observed that some developers mistakenly exacerbate API compatibility issues by incorrectly modifying compatibleSdkVersion. For example, in a particular update, the applications\_contacts application in the main openharmony organization reduced the compatibleSdkVersion to 9, resulting in the emergence of new compatibility issues from the invocation of certain APIs introduced in API 10. These observations show that most OpenHarmony developers currently pay insufficient attention to API compatibility issues and lack effective methods to identify and fix them.

\subsubsection{Fix Recommendation}  We further explored targeted repair strategies for specific compatibility problems. A concrete example involves the \textit{MultiMedia} application, which exhibited forward compatibility issues due to the use of the removed API mediaLibrary.getMediaLibrary in ohos.multimedia.mediaLibrary. Listing~\ref{code:multimedia_fix} illustrates this case. The API declaration, extracted from the file ``@ohos.multimedia.mediaLibrary.d.ts'', indicates that getMediaLibrary was introduced in API version 6, deprecated in API version 9, and removed in API version 12, resulting in a lifecycle of [6,11]. The ``@useinstead'' annotation suggests replacing it with an API from the ``ohos.file.picker'' module. By consulting the OpenHarmony documentation\cite{ohos_medialibrary}, we identified the replacement API as ``photoAccessHelper.getPhotoAccessHelper'' from module ``SystemCapability.FileManagement.PhotoAccessHelper.Core''. The problematic code in MultiMedia used getMediaLibrary to obtain a media library instance. To resolve the compatibility issue, we replaced this call with photoAccessHelper.getPhotoAccessHelper, ensuring compatibility with API versions 9 and above. After the fix, \tool{} detected no issues for this API, and runtime tests on API 12 devices confirmed stability.

\begin{lstlisting}[language=TypeScript, basicstyle=\ttfamily\scriptsize, xleftmargin=1em, label={code:multimedia_fix}, caption={Fix Recommendation for MultiMedia Application}]
// @ohos.multimedia.mediaLibrary.d.ts
declare namespace mediaLibrary {
  /**
   * Obtains a MediaLibrary instance.
   * @since 6
   * @syscap SystemCapability.Multimedia.MediaLibrary.Core
   * @deprecated since 9
   * @useinstead ohos.file.picker
   */
  function getMediaLibrary(): MediaLibrary;
}
// Original problematic code
const media = mediaLibrary.getMediaLibrary(context);
// Fixed code
const media = photoAccessHelper.getPhotoAccessHelper(context);
or
if (sdkAPIVersion < 9) {
    media = mediaLibrary.getMediaLibrary(uiAbility.context);
}
\end{lstlisting}


\find{\textbf{Answers to RQ3:}  
1. Most OpenHarmony developers are either unaware of existing compatibility issues or do not treat them as priorities. 2. Effective compatibility needs understanding API lifecycles and applying semantically equivalent replacements where necessary. 
}

\section{Related work}
\label{sec:relatedwork}

Since OpenHarmony is still a relatively new system with limited investigation, we turn to prior research in the Android ecosystem and other relevant topics for valuable insights~\cite{wang2025multi}. The problem of compatibility issues has persistently posed challenges for both users and developers, making it a prominent and extensively studied subject.


\textbf{API Evolution.}
As APIs continue to evolve, developers and professionals face numerous challenges in software maintenance~\cite{martin2016survey, li2023software}. In the literature, various studies on API evolutions across different ecosystems have been published, with the aim of facilitating and advancing the process of managing API changes.
Sawant et al.~\cite{sawant2016reaction, sawant2018reaction} investigated the consequences of deprecating five Java APIs within a sample of 25,357 customers. Following this, they conducted semi-structured interviews (2015) with 17 third-party Java API providers and surveyed 170 Java developers to gain a comprehensive understanding of the requirements and perspectives surrounding deprecation, encompassing both API producers and users~\cite{sawant2018understanding}.
Li and his colleagues~\cite{li2016investigation, li2018characterising, li2020cda} have investigated the ripple effects caused by API evolution within the Android framework. 
There are works to investigate the evolution of Python libraries and reveal that there are compatibility issues~\cite{zhang2020python, wang2020exploring, quan2022characterizing}. For example, Wang et al.~\cite{wang2020exploring} proposed dlocator to locate the usage of deprecated APIs in client applications.
The authors further proposed SnifferDog~\cite{wang2021restoring} to automatically restore the execution environment of Jupyter notebooks based on the analysis of API usages and the pre-build API bank.
Although previous studies have focused predominantly on evaluating the implications of API evolution within diverse software ecosystems, this aspect has yet to be investigated within OpenHarmony applications. Our research not only examines compatibility concerns that stem from the evolution of the OpenHarmony SDK, but also introduces a tool called \tool{} for the detection and resolution of such issues in OpenHarmony applications.

\textbf{Compatibility Issues.}
Over the past few years, researchers have proposed a range of tools aimed at identifying Android compatibility issues~\cite{cai2019large}. 
Dynamic-based approaches~\cite{fazzini2017automated, ki2019mimic} have been adopted to generate tests to identify app GUI inconsistencies across Android devices. However, these approaches often rely on random test generation strategies, which can be ineffective in triggering inconsistencies within the vast search space of the configuration attributes of an app. 
Most tools~\cite{wei2016taming, huang2018understanding, yang2018android, he2018understanding, li2018cid, mahmud2021android, scalabrino2019data} adopted static-based approaches for detecting compatibility issues by their predefined patterns to facilitate detection. 
However, it is necessary to continue to adapt those predefined patterns associated with compatibility issue detection tools to support the latest changes in the mobile ecosystem~\cite{liu2022automatically, chen2024your}. 
Scalabrino et al.~\cite{scalabrino2019data, scalabrino2020api} conclude the rules through an analysis of conditional API usage and assign a confidence level to each rule, thus facilitating the identification of potentially problematic APIs for Android compatibility issues. 
Li et al.~\cite{li2018cid} achieve detection of Android compatibility issues by constructing a model of the API lifecycle and conducting comparisons of API calls to assess whether an API is responsible for compatibility concerns.
In communities outside of the Android ecosystem, there is notable research focused on compatibility concerns. Ma et al.~\cite{ma2023cid4hmos} employ a similar approach to CiD, using HarmonyOS Java SDKs to examine the development of HarmonyOS Java APIs. Zhou et al.~\cite{zhou2016api} introduced a prototype tool called Deprecation Watcher, which demonstrates high precision and recall in identifying the usage of deprecated APIs within the Java source code.
Although there have been numerous studies on compatibility issues, our research stands out as pioneering the exploration of compatibility issues within applications developed using the newly designed language, ArkTS, for the innovative OpenHarmony OS.

\section{Conclusion}
\label{sec:conclusion}
OpenHarmony's rapidly evolving APIs create compatibility risks that Android-focused tools cannot adequately address. We presented \tool{}, the first automated methodology for detecting API-related compatibility issues in OpenHarmony applications. Applied to OpenHarmony community apps, \tool{} identified and categorized recurring issues and produced repair recommendations. The results demonstrate their prevalence and the need for lifecycle-aware remediation. The tool, replication package, and experimental data are available at \url{https://github.com/ElouanDai/HapCiD}.

\bibliographystyle{ACM-Reference-Format}
\bibliography{myBibliography}

\end{document}